\documentclass[prb,twocolumn,superscriptaddress,showpacs,floatfix,floats]{revtex4-1}
\usepackage{textcomp,graphicx,natbib}
\newcommand{\invcm}{\mbox{cm}\ensuremath{^{-1}}}

\newcommand{\dgC}{\ensuremath{^\circ\mbox{C}}}

\begin{document}
\title{Possible coupling between magnons and phonons in multiferroic $\rm CaMn_7O_{12}$}
\affiliation{Institute of Physics, Academy of Sciences of the Czech Republic, Na Slovance 2, 182 21 Prague~8, Czech Republic}
\affiliation{ISIS Facility, Rutherford Appleton Laboratory, Chilton, Didcot, United Kingdom}
\affiliation{Institut Laue-Langevin, Bo\^\i te Postale 156, 38042 Grenoble Cedex 9, France}
\affiliation{Department of Condensed Matter Physics, Faculty of Mathematics and Physics, Charles University, Ke Karlovu 5, 121 16 Prague 2, Czech Republic}
\affiliation{Grenoble High Magnetic Field Laboratory, CNRS - 25, avenue des Martyrs, Grenoble Cedex 9, France}
\author{Filip Kadlec}
\thanks{Authors to whom correspondence should be addressed}
\email[e-mail: ]{kadlecf@fzu.cz; kamba@fzu.cz}
\affiliation{Institute of Physics, Academy of Sciences of the Czech Republic, Na Slovance 2, 182 21 Prague~8, Czech Republic}
\author{Veronica Goian}
\affiliation{Institute of Physics, Academy of Sciences of the Czech Republic, Na Slovance 2, 182 21 Prague~8, Czech Republic}
\author{Christelle Kadlec}
\affiliation{Institute of Physics, Academy of Sciences of the Czech Republic, Na Slovance 2, 182 21 Prague~8, Czech Republic}
\author{Martin Kempa}
\affiliation{Institute of Physics, Academy of Sciences of the Czech Republic, Na Slovance 2, 182 21 Prague~8, Czech Republic}
\author{P\v remysl Van\v ek}
\affiliation{Institute of Physics, Academy of Sciences of the Czech Republic, Na Slovance 2, 182 21 Prague~8, Czech Republic}
\author{Jon Taylor}
\affiliation{ISIS Facility, Rutherford Appleton Laboratory, Chilton, Didcot, United Kingdom}
\author{St\'ephane Rols}
\affiliation{Institut Laue-Langevin, Bo\^\i te Postale 156, 38042 Grenoble Cedex 9, France}
\author{Jan Prokle\v{s}ka}
\affiliation{Department of Condensed Matter Physics, Faculty of Mathematics and Physics, Charles University, Ke Karlovu 5, 121 16 Prague 2, Czech Republic}
\author{Milan Orlita}
\affiliation{Grenoble High Magnetic Field Laboratory, CNRS - 25, avenue des Martyrs, Grenoble Cedex 9, France}
\author{Stanislav Kamba}
\affiliation{Institute of Physics, Academy of Sciences of the Czech Republic, Na Slovance 2, 182 21 Prague~8, Czech Republic}

\begin{abstract}
Spin and lattice dynamics of CaMn$_{7}$O$_{12}$ ceramics were investigated using 
infrared, THz and inelastic neutron scattering (INS) spectroscopies in 
the temperature range 2 to 590\,K, and, at low temperatures, in applied 
magnetic fields of up to 12\,T. On cooling, we observed phonon splitting 
accompanying the structural phase transition at $T_{\rm C}$ = 450\,K as well as the 
onset of the incommensurately modulated structure at 250\,K.
In the two antiferromagnetic 
phases below $T_{\rm N1} = 90\,\rm K$ and $T_{\rm N2} = 48\,\rm K$, 
several infrared-active excitations emerge in the meV range; their 
frequencies correspond to the maxima in the magnon density of states obtained by 
INS. At the magnetic phase transitions,  these modes display strong 
anomalies and for some of them, a transfer of dielectric strength from the 
higher-frequency phonons is observed.  We propose that these modes 
are electromagnons. Remarkably, at least two of these modes remain active also in the 
paramagnetic phase; for this reason, we call them paraelectromagnons.
In accordance with this observation, quasielastic neutron scattering revealed 
short-range magnetic correlations persisting within temperatures up to 
500\,K above $T_{\rm N1}$.

\end{abstract}
\date{\today}

\maketitle
\section{Introduction}
Electromagnons are collective spin excitations occurring in many
magnetoelectric multiferroics, contributing to their dielectric response
and potentially also to the magnetic 
one\cite{Pimenov08,Kida09,Shuvaev11,Kida12}. They appear frequently in the 
THz-range spectra below phonons, and they are frequently understood as spin 
waves excited by the electric component of the incident electromagnetic 
radiation; more precisely, they contribute to the off-diagonal components of the 
magneto-electric susceptibility tensor.  Although electromagnons were observed 
in a variety of multiferroics, there is still no universal microscopic theory 
which would explain the mechanisms giving rise to electromagnons.  Depending on 
the crystalline and magnetic structure, the magnons can be activated in the 
THz-range dielectric spectra by one of the following magnetoelectric coupling 
mechanisms: the magnetostriction \cite{ValdesAguilar09,Stenberg12}, inverse 
Dzyaloshinskii-Moriya interaction \cite{Katsura07,Mochizuki11} or spin-dependent 
hybridization of the $p$ and $d$ orbitals \cite{Seki10}.  Interestingly, in one 
material, the magnetoelectric coupling mechanism leading to activation of 
electromagnons can be different from the one giving rise to the static 
spin-induced polarization.  For example, in the well-known multiferroic $\rm 
TbMnO_3$, the polarization is induced by the inverse Dzyaloshinskii-Moriya 
mechanism, while at least some electromagnons are activated by magnetostriction 
\cite{ValdesAguilar09}.  Furthermore, it is still unclear how the electromagnons 
exhibiting a wavevector off the Brillouin-zone (BZ) center can be excited by the 
incident photons with wavevectors near the BZ center, $q \approx 0$; it seems 
that this may be allowed by the modulation of the magnetic structure.  Generally 
speaking, the electromagnons as excitations which are both electric- and 
magnetic-dipole active reveal the dynamical features of the magnetoelectric 
coupling \cite{ValdesAguilar09,Stenberg12,Katsura07,Mochizuki11,Seki10}, and 
studying their temperature- and magnetic-field-dependent properties may provide 
insight into the symmetry of the spin subsystem of the materials.

Among the many spin-driven multiferroics which have been studied intensively in 
the last years, the mixed-valence manganate $\rm CaMn_7O_{12}$ distinguishes 
itself by several prominent features. At low temperatures, the material exhibits 
a strong magnetoelectric \cite{Zhang11,Sanchez-Andujar09} and magnetoelastic 
\cite{Slawinski10} coupling.  The spontaneous ferroelectric 
polarization saturates at $P_{\rm s}=2.9\rm\,mC/m^{2}$ below the temperature of 
40\,K, which represents the highest value among the currently known magnetically driven 
ferroelectrics \cite{Johnson12}. As a consequence of the magnetoelectric coupling, the 
polarization decreases in external magnetic field \cite{Zhang11}. The spins of 
the Mn ions are arranged in a way giving rise to a so-called ferroaxial 
ordering, corresponding to a spiral arrangement of the spins with a chiral 
symmetry \cite{Johnson12}. In different studies, all the three interactions 
cited above were identified as the dominant coupling mechanism leading to the 
exceptionally high electric polarization
\cite{Johnson12,Zhang13,Lu12,Perks12}.  There is no clear agreement about 
the respective contributions of these mechanisms as yet. Moreover, the 
record value of $P_{\rm s}$ for a multiferroic material requires an 
explanation, which appears to be a challenging task.

The usual way of identifying electromagnons consists in measuring the 
polarization-sensitive spectra of the infrared (IR)-active excitations polarized 
along all the crystallographic axes \cite{Pimenov08,Kida09,Takahashi08}. In 
this 
case, selection rules for the polarization of both the electric and 
the magnetic vectors of the electromagnetic radiation must be taken into 
account. Electromagnons can be also distinguished from phonons or purely 
magnetic active spin waves (magnons) using measurements of the so-called directional 
dichroism \cite{Kezsmarki11,Kezsmarki14,Takahashi12}. However, both these approaches fail if sufficiently large crystals 
(thin slabs with two faces of typically several $\rm mm^2$) are not available. 
We have shown recently \cite{Kadlec13} that in such a case, the  electromagnons can be recognized using a combination of 
unpolarized IR spectra and inelastic neutron scattering (INS).  If 
there are IR-active excitations activated below a magnetic phase 
transition temperature, which receive dielectric strength from polar phonons, 
and if these excitations are observed as magnons in the INS spectra, then these 
excitations can be attributed to electromagnons. In view of this method we have 
studied, using Fourier-transform IR (FTIR), time-domain THz and INS 
spectroscopies, the low-energy spectra of polycrystalline $\rm CaMn_7O_{12}$, in 
a broad temperature interval from 2 to 590\,K.

\section{Properties of $\rm CaMn_7O_{12}$}
At high temperatures, $\rm CaMn_7O_{12}$ has a cubic symmetry with the space 
group $Im\overline 3$. The atomic arrangement can be considered \cite{Zhang13} as a 
variant of the perovskite structure with a set of regularly alternating four 
$\rm ABO_{3}$ units and a sum formula $\rm (CaMn_3)Mn_4O_{12}$. This phase 
features two different
Mn positions; one occupying the oxygen octahedra centers with a mixed valence of 
3.25, and one placed in the perovskite cube
corners, exhibiting a  valence of 3 \cite{Przenioslo02}. At $T_{\rm 
CO}= 440\,\rm K$, the compound undergoes a charge-ordering metal-insulator 
\cite{Troyanchuk98} phase transition to a phase with a rhombohedral structure 
(space group $R\overline 3$)\cite{Bochu80}. In this phase, the mixed-valence Mn 
atoms shift into two Wyckoff positions---9d and 3b, displaying valences of 
+3 and +4, respectively; the two kinds of octahedra are subject to two 
different Jahn-Teller distortions\cite{Bochu80}.

The next phase transition at $T_{\rm inc}=250\,\rm K$ gives 
rise to Mn orbital ordering \cite{Perks12}. At the same time, an 
incommensurate structural modulation along the $c$-axis \cite{Slawinski12a} with 
a propagation vector $q_{\rm c}\approx(0,0,2.08)$ appears \cite{Perks12}. 
On further cooling, the crystal undergoes two magnetic phase transitions. At 
$T_{\rm N1}=90\,\rm K$, a long-range helical antiferromagnetic order 
with a magnetic superspace group $R31^{\prime}(00\gamma)ts$ 
(Ref.~\onlinecite{Slawinski12}) and a 
spin-induced ferroelectric polarization develop 
\cite{Zhang11,Johnson12}; the magnetic propagation vector 
$q_{\rm m}=q_{\rm c}/2$ is temperature-independent \cite{Slawinski10}. The orbital modulation was identified as one of the 
key elements stabilizing the helical magnetic structure \cite{Perks12}. Below 
$T_{\rm N2}=48\,\rm K$, a different antiferromagnetic phase is observed, where 
the magnetic ordering can be described by two temperature-dependent propagation 
vectors $q_{\rm m1}$ and $q_{\rm m2}$ \cite{Slawinski12,Johnson12}. These are
parallel with the $c$-axis, and their average is equal to $q_{\rm m}$. Near $T_{\rm N2}$, a temperature hysteresis of the 
magnetization was detected.\cite{Sannigrahi13} It was  shown that the magnetization $M$ below $T_{\rm N2}$ exhibits a 
memory of the magnetic field applied during the cool-down. The latter 
observation was interpreted in terms of superparamagnetic clusters, which could 
be also at the origin of the $M(T)$ hysteresis\cite{Sannigrahi13}.

\section{Samples and experimental details}
According to earlier published works, it is very difficult to produce single 
crystals substantially larger than a fraction of a cubic millimeter,  
\cite{Johnson12} which would be necessary for spectroscopic studies with 
incident waves polarized along the crystallographic axes. Therefore, such 
studies must be performed on polycrystalline or powder samples only.
Our $\rm CaMn_7O_{12}$ powder was prepared by the modified Pechini method using 
calcium and manganese nitrates in a stoichiometric ratio, ethylene glycol and 
water. The solution was slowly heated up to 250\,\dgC{}, the resulting black foam was 
ground and the powder was annealed at 830\,\dgC{} (24\,h), 930\,\dgC{} (48 h) 
and 950\,\dgC{} (48 h) with intermittent grinding. Subsequently, the powder was 
uniaxially pressed at 400\,MPa (3 min) to a pellet with a diameter of 12\,mm and 
sintered at 950\,\dgC{} for 48 h. The density of the pellet was about 80\% of 
the single crystal value. Its phase purity was verified by means of powder x-ray 
diffraction, excluding other stoichiometry than $\rm CaMn_7O_{12}$ 
beyond one weight per cent. The pellet was thinned down to a thickness of 0.36\,mm 
and optically polished.

IR reflectance and transmittance measurements with a resolution of 2 and 
0.5\,\invcm{}, respectively, were performed using the FTIR spectrometer Bruker 
IFS-113v. An Oxford Instruments Optistat optical cryostat with polyethylene 
windows was used for sample cooling down to 10\,K, and a liquid-He-cooled Si 
bolometer operating at 1.6\,K served as a detector. For measurements at high 
temperatures, a high-temperature cell (SPECAC P/N 5850) was used. Far-IR 
transmittance with an applied magnetic field was measured using another Bruker 
IFS-113v spectrometer and a custom-made cryostat with a superconducting magnet.  
This setup allowed measurements at 2 and 4\,K in the Faraday geometry with an 
external magnetic field of up to $B_{\rm ext}=12\,\rm T$. Time-domain THz 
spectroscopy was performed by measuring the complex sample transmittance using 
custom-made spectrometers based on Ti:sapphire femtosecond lasers; one with an 
Optistat cryostat, used for measurements without magnetic field, and one with an 
Oxford Instruments Spectromag cryostat, enabling measurements with $B_{\rm 
ext}\leq7\,\rm T$. Here, the Voigt configuration was used; the electric 
component of the THz radiation $\mathbf{E}_{\rm THz}$ was set parallel to 
$\mathbf{B}_{\rm ext}$. Reference measurements with an empty aperture, 
enabling a reliable determination of the transmittance, were performed 
systematically at each value of applied magnetic field. The same results, 
within the experimental accuracy, were obtained also for $\mathbf{B}_{\rm 
ext}\perp \mathbf{E}_{\rm THz}$.

INS experiments using about 3\,g of loose $\rm CaMn_7O_{12}$ powder were 
performed first at 10\,K in the IN4 time-of-flight spectrometer at the Institut 
Laue-Langevin (ILL Grenoble, France), and within temperatures from 10 to 590\,K 
using the MARI time-of-flight spectrometer at the ISIS facility, Rutherford Appleton 
Laboratory (UK). The low-temperature dielectric permittivity 
	at 1\,kHz was measured in magnetic field of up to 9\,T using the high 
precision capacitance bridge Andeen-Hagerling 2500A in a PPMS 9T instrument, Quantum design.

\section{Results and discussion}
	\subsection{IR spectroscopy}
\begin{figure}
	  \centering
	  \includegraphics[width=0.9\columnwidth]{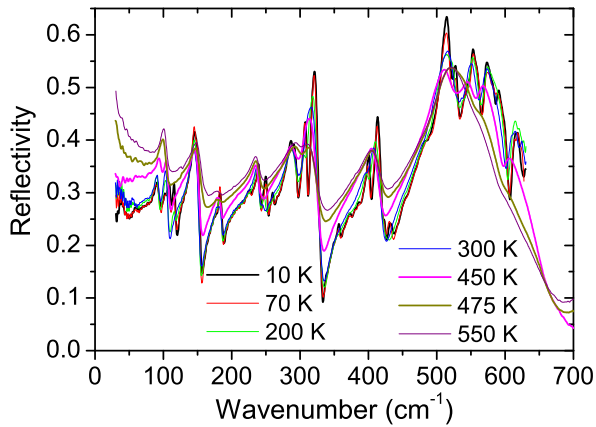}
	  \caption{IR reflectivity spectra of $\rm CaMn_7O_{12}$ ceramics in a broad 
	  temperature range. The low-temperature spectra above 620\,\invcm{} could 
  not be obtained due to opacity of the polyethylene windows of the cryostat.}
	  \label{fig:IRrefl}
  \end{figure}
\begin{figure}
	  \centering%
	  \includegraphics[width=\columnwidth]{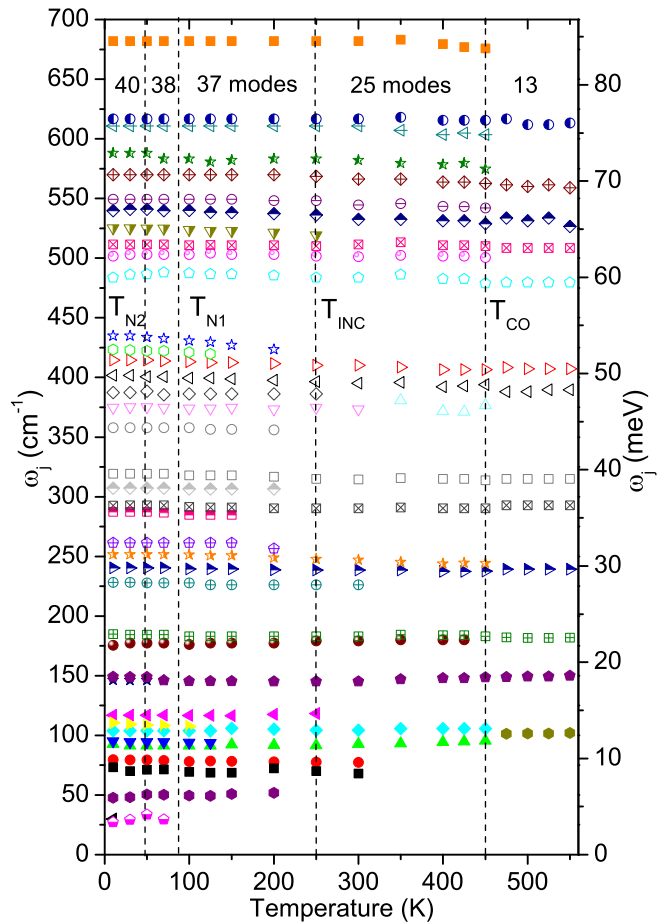}
	  \caption{Temperature dependence of the mode eigenfrequencies obtained 
	  from fits of IR and THz spectra. The numbers of modes observed in 
  different phases are stated above.}
	  \label{fig:mody-T}
  \end{figure}

Fig.\ \ref{fig:IRrefl} shows the IR reflectivity spectra of $\rm 
CaMn_7O_{12}$ ceramics at selected temperatures in all known phases. In the 
high-temperature cubic phase, 13 modes are observed. Moreover, below ca.\ 
150\,\invcm, an increase in reflectivity towards low frequencies is seen, 
typical of absorption on free carriers. Below $T_{\rm CO}$, the sample becomes 
insulating and the reflection on the free carriers disappears. Upon decreasing temperature 
and lowering of crystal
symmetry, the number of observed polar phonons increases and at the 
lowest temperatures, 40 IR-active modes can be distinguished in the 
reflectivity and transmittance spectra. 
The  frequencies of all observed modes in the whole temperature range are 
shown in Fig.~\ref{fig:mody-T}. Here and in the following, $\omega_j$ 
stands for the mode eigenfrequency of the $j$-th mode which can be, in principle, either the 
transverse optical phonon frequency $\omega_{\rm TO}$ or an (electro)magnon frequency 
$\omega_{\rm mag}$.

We have performed a factor-group analysis of BZ-center lattice vibrations for the cubic $Im\overline{3}$ and trigonal $R\overline{3}$ 
symmetries. In both crystal structures, the primitive unit cell contains a 
single formula unit with 20 atoms, implying 60 vibrational degrees of
freedom. For the cubic structure, we obtained the following phonon counts:
	\begin{eqnarray}
	\label{cubic-anal}
	\nonumber \Gamma_{Im\overline{3}}&=12F_{u}(\mbox{IR})+2A_{u}(-)+2E_{u}(-)+\\
		&+2A_{g}(\mbox{R})+ 2E_{g}(\mbox{R})+4F_{g}(\mbox{R}).
	\end{eqnarray}
where IR, Raman-active (R) and silent modes (--) are marked in the parentheses. 
The analysis for the trigonal structure yields:
	\begin{equation}
	\label{trigonal-anal}
	\nonumber \Gamma_{R\overline{3}}=14E_{u}(\mbox{IR})+14A_{u}(\mbox{IR})+6A_{g}(\mbox{R})+6E_{g}(\mbox{R}).
	\end{equation}
Thus, after subtraction of acoustic modes, 11 and 26 IR-active modes are
expected in the cubic and trigonal phases, respectively.

\begin{figure}
	  \centering
	  \includegraphics[width=0.9\columnwidth]{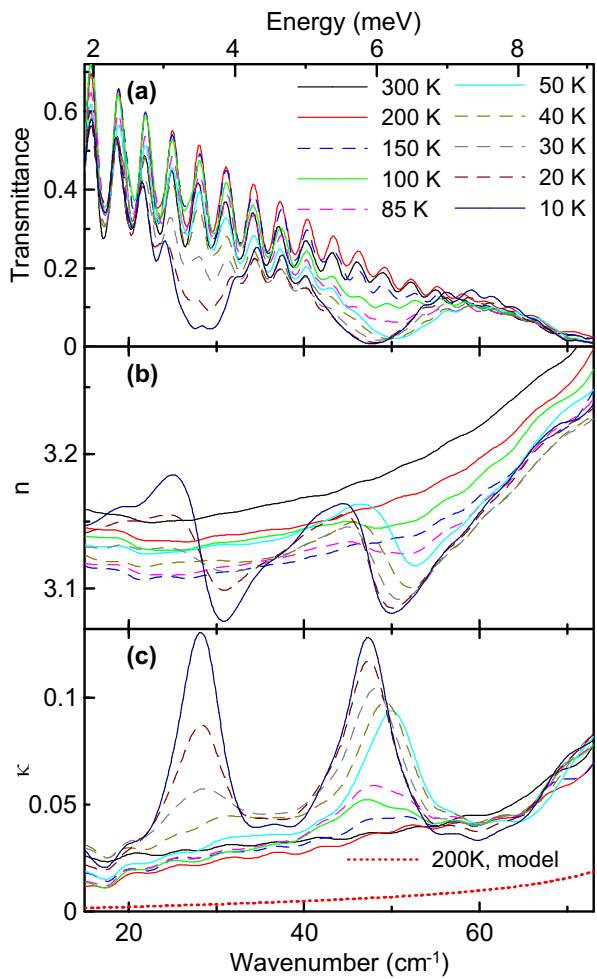}
	  \caption{Spectra of (a) far-IR transmittance, (b) real and (c) 
		  imaginary parts of the complex refractive index $N=n+ 
			  \mbox{i}\kappa$ obtained by THz 
	  time-domain spectroscopy. The red dotted line in part (c) shows
	  the low-frequency wing from the phonons at frequencies above 90\,\invcm{},
	  obtained from fits.}
	  \label{fig:lowFallT}
\end{figure}

The most interesting changes related to the magnetic ordering occur in the 
frequency range below 80\,\invcm{} (10\,meV); they can be well observed in both the FTIR and 
the THz transmittance spectra. Whereas the FTIR transmittance offers a 
	higher sensitivity especially for photon energies above 7\,meV and a better 
	spectral resolution, the measured absorption lines are overlapping with the 
	interference patterns (i.e., the periodic oscillations seen in 
		Fig.~\ref{fig:lowFallT}a) 
	arising from multiple reflections of the far-IR waves on the parallel sample 
surfaces. By contrast, the THz time-domain spectroscopy directly 
	reveals the complex optical constants up to 9\,meV, it avoids the 
	interferences by separating the multiple reflections in time, but it 
is difficult to achieve a good resolution of narrow spectral lines. Thus, using 
both these complementary techniques is the best approach for obtaining a 
detailed spectral information in the far-IR range.

Fig.~\ref{fig:lowFallT}a shows the far-IR transmittance as a 
function of temperature. Upon cooling, between 300 and 
200\,K, the transmittance slightly increases due to a decrease in the phonon 
damping and the related losses. In contrast, below 200\,K, a decrease in the 
transmittance occurs. At first, a broad absorption band around 50\,\invcm{} 
gradually develops with cooling, with the fastest increase in its intensity near 
$T_{\rm N1}$ where the band becomes narrower. Mainly below $T_{\rm N2}$, another 
absorption band is seen near 28\,\invcm. Analogous changes occur in the spectra 
of the  complex refractive index obtained by THz spectroscopy (see 
Figs.~\ref{fig:lowFallT}b, c). 

Both FTIR transmittance and THz spectra were fitted assuming a sum of 
Lorentz oscillators describing a dispersion in the dielectric permittivity 
only. We are keenly aware that this is a rough simplification. Whereas this 
model is perfectly valid for phonons, i.e.\ excitations located approximately 
beyond 90\,\invcm{}, the lower-frequency ones can contribute, to some extent, 
also to the magnetic permeability. Furthermore, the model of a sum of 
harmonic Lorentz oscillators is not exact in magneto-electric multiferroics, 
since the off-diagonal components of the magneto-electric susceptibility 
tensor contribute to the complex index of refraction \cite{Kezsmarki14}. 
Nevertheless, we used the classical oscillator model to describe the 
dispersion in $N$, similarly to earlier works (see, e.g., Ref.  
\onlinecite{Kida09}), which allowed us to determine quantitatively the 
parameters of all observed excitations.

\begin{figure}
	  \centering
	  \includegraphics[width=\columnwidth]{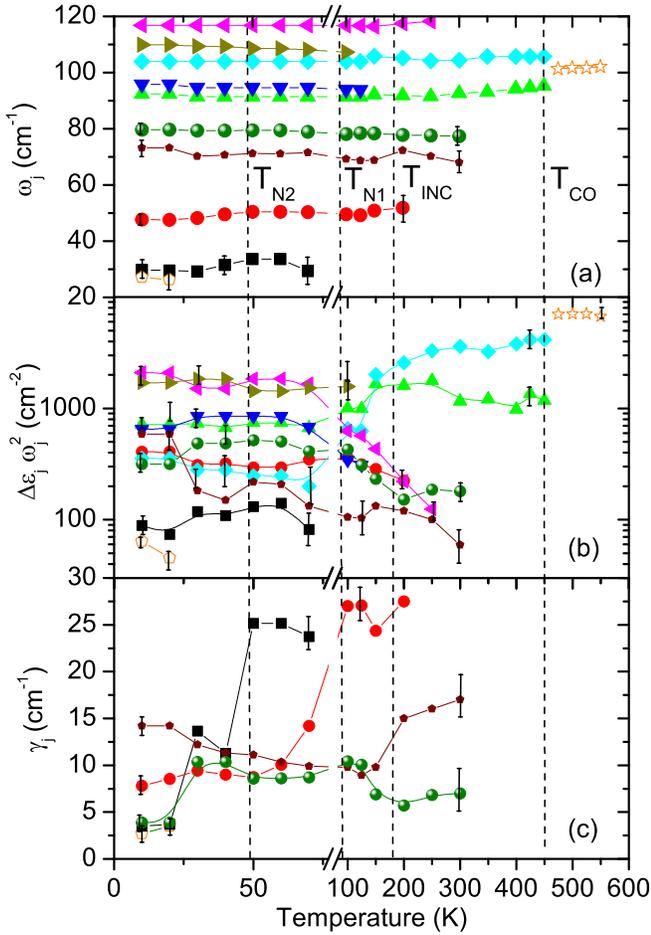}
	  \caption{Temperature dependence of (a) frequencies $\omega_j$ of the 
	  vibrational modes observed below 120\,\invcm{}, (b) their 
	  oscillator strengths, and (c) dampings of excitations below 
	  80\,\invcm{}. These parameters were obtained by fitting the measured 
	  FTIR and THz spectra. For clarity, the error bars are shown only 
	  at selected temperatures. The errors in frequencies of modes above 
	  90\,\invcm{} are smaller than the symbols.}
	  \label{fig:lowf-modes}
  \end{figure}

Careful simultaneous fitting of the IR transmittance, FTIR reflectivity and THz 
spectra implies that while there is a 
single absorption mode near 50\,\invcm, the shape of the spectra near 30 and 
75\,\invcm{} is such that for their correct modelling, a doublet for each of 
these regions is required. The temperature dependence of the mode frequencies is 
shown in Fig.~\ref{fig:lowf-modes}a.  The marked anomalies in 
the
intensities (i.e., oscillator strengths ${\Delta\varepsilon_j}\,\omega_j^2$) 
and damping constants $\gamma_j$ of 
the modes below 60\,\invcm{} near $T_{\rm N1}$ and $T_{\rm N2}$ suggest that 
these are connected with the magnetic order (see 
Fig.~\ref{fig:lowf-modes}b).  
Interestingly, the mode seen near 50\,\invcm{} is clearly present with a 
high damping also in the 
spectra in the paramagnetic phase at least up to 200\,K. As for the doublet near 
75\,\invcm{}, the sensitivity of our experimental methods is somewhat lower in 
this region, but the two modes can be clearly discerned at 10\,K. Even at higher 
temperatures, absorption at these frequencies is necessary for a correct 
description of the spectra by fitting. As an example, Fig.~\ref{fig:lowFallT}c 
shows a spectrum of the extinction coefficient calculated using a model where 
only the harmonic oscillators with eigenfrequencies $\omega_j>90\,\invcm{}$ were 
included; omitting the modes at and below 80\,\invcm{} leads to a marked 
discrepancy from the experimental data. Thus, thanks to modelling the spectra, 
we were able to follow these modes between 10 and 300\,K. Above  room 
temperature, they may still exist, but the sample becomes opaque in this 
spectral range, and the modes are probably too weak to be seen in the 
reflectivity spectra.

\begin{figure}
      \centering
      \includegraphics[width=0.9\columnwidth]{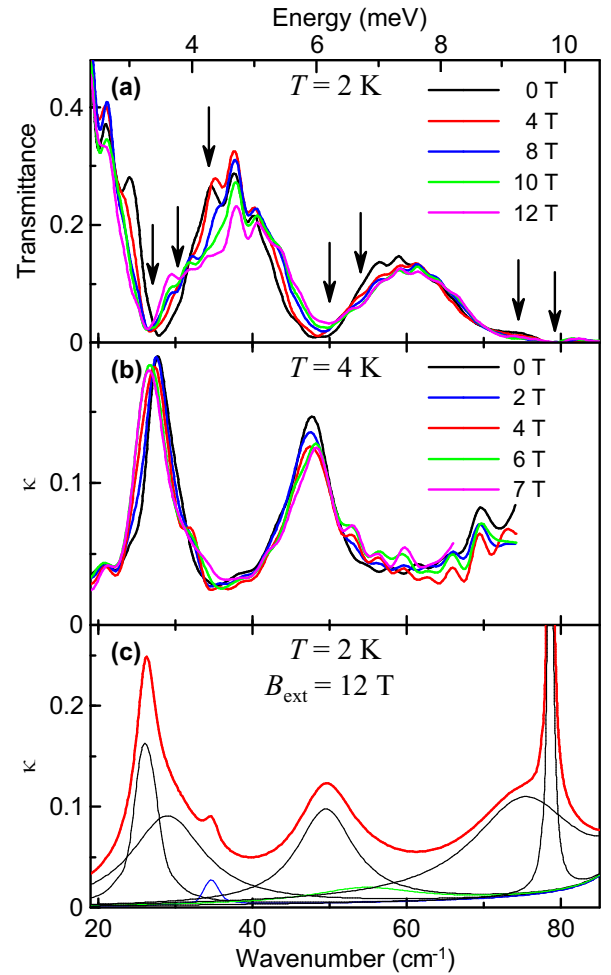}
	  \caption{(a) Low-energy FTIR transmittance in external magnetic field 
	  (Faraday geometry)
	  at $T=2$\,K. The arrows mark frequencies of modes at 
		  $B_{\rm ext}=12\,\mbox{T}$. (b) Extinction coefficient at $T=4$\,K as a function of $B_{\rm ext}$, 
		  obtained by THz spectroscopy in the Voigt geometry. (c) 
		  Contributions of individual modes to the extinction coefficient at  
	  $T=4$\,K and $B_{\rm ext}=12$\,T, obtained from the fit of the FTIR 
	  transmittance.}
      \label{fig:FIR-B}
  \end{figure}

\begin{figure}
	  \centering
	  \includegraphics[width=\columnwidth]{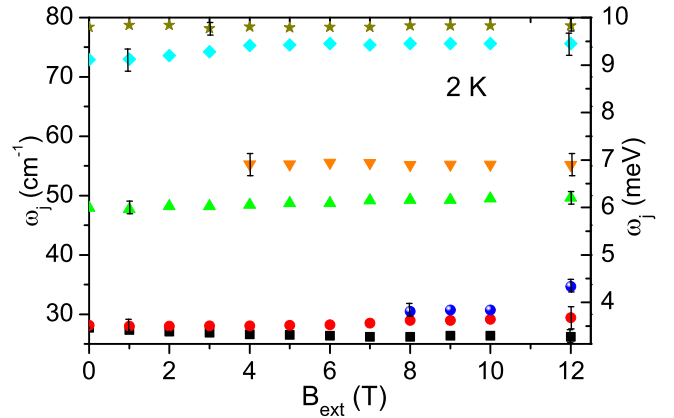}
	  \caption{Magnetic-field dependence of the mode frequencies obtained from 
		  the spectra fits for $T=2$\,K.}
	  \label{fig:freq-4K-B}
  \end{figure}
  
	  Fig.~\ref{fig:FIR-B} shows the spectra of IR transmittance and of the 
	  extinction coefficient measured by THz spectroscopy below 80\,\invcm{} at the lowest 
	  temperatures, as a function of $B_{\rm ext}$. The magnetic-field dependence of 
	  the mode frequencies is shown in Fig.~\ref{fig:freq-4K-B}.
	  Two kinds of changes were observed with increasing $B_{\rm ext}$. First, the 
	  frequencies of the modes shift slightly; if a field of $B_{\rm ext}=7$\,T is 
	  applied, the modes at 48\,\invcm{} and 28\,\invcm{} harden by 
	  about 1\,\invcm{}, whereas the one near 27\,\invcm{} softens by about 
	  1.5\,\invcm{}. Second, additional weaker but clearly observable satellites arise 
	  progressively around 55 and 30\,\invcm{} (see Fig.\ \ref{fig:FIR-B}). Whereas the position of the satellite near 55\,\invcm{} is 
	  independent of the magnetic field, the splitting of the one appearing near 
	  30\,\invcm{}  increases up to ca.~6\,\invcm{} for $B_{\rm ext}=12$\,T. 
	  The shapes of the individual lines, contributing to the  
	  absorption at this value of $B_{\rm ext}$, are shown in 
	  Fig.~\ref{fig:FIR-B}c. Such a splitting 
	  corresponds qualitatively to a behavior which was theoretically predicted 
	  \cite{Chen11} for multiferroics with a spiral arrangement of spins.

	The absorption peaks observed for energies below 10\,meV can, in principle, 
	be due to phonons or magnons; the changes in the spectra with temperature 
	and
	magnetic field provide indications for either of these types. The doublet 
	 at 75 and 78\,\invcm{} (Fig.~\ref{fig:freq-4K-B}) is likely due 
	to polar phonons, because it is active already at 300\,K. However, one 
	of these modes can be also partly due to a spin excitation, since the magnon 
	density of states (DOS) exhibits a broad maximum in this energy range (see the discussion 
of the INS data below). In the rhombohedral phase, 25 modes (including these two) 
are observed, while the crystal symmetry allows up to 26 zone-center phonons (see 
Eq.~\ref{trigonal-anal}). The missing mode can be weak or overlapping with another one. At $T_{\rm inc}$, $\rm CaMn_7O_{12}$ undergoes a 
structural phase transition to an
incommensurate structure, where 12 new modes appear in the spectra; this 
	kind of transitions is known to activate polar phonons due to folding of the 
	BZ \cite{Petzelt81}.  Among them, there are the modes below 60\,\invcm{} 
	which we attribute to spin waves owing to the significant changes in 
	intensities and dampings they exhibit near the magnetic phase transitions and splitting 
	in external magnetic field.
	Further confirmation of the magnetic character of these modes was obtained 
	by INS.
	
	The oscillator strengths of the excitations below 
	80\,\invcm{} increase below $T_{\rm N1}$ and $T_{\rm N2}$, whereas some 
	phonon oscillator strengths decrease below these critical temperatures (see 
	Fig.~\ref{fig:lowf-modes}b). Some changes in the
	oscillator strengths can be caused by mode splitting near 100\,K which 
	clearly occurs for phonons above 95\,\invcm{}. At the same time, it is very 
	likely that some strength is also transfered to the newly appearing 
	excitations, which must have a magnetic nature, as 
	discussed above.  This would be a signature of dynamic 
	magneto-electric coupling.  Therefore, the observed low-frequency 
	excitations are likely to be electromagnons. A definite confirmation of the 
	electromagnons could be obtained only by an analysis of polarized THz and IR 
	spectra of single crystals, which are currently not available.

\subsection{Inelastic neutron scattering}
\begin{figure}
  \centering
  \includegraphics[width=\columnwidth]{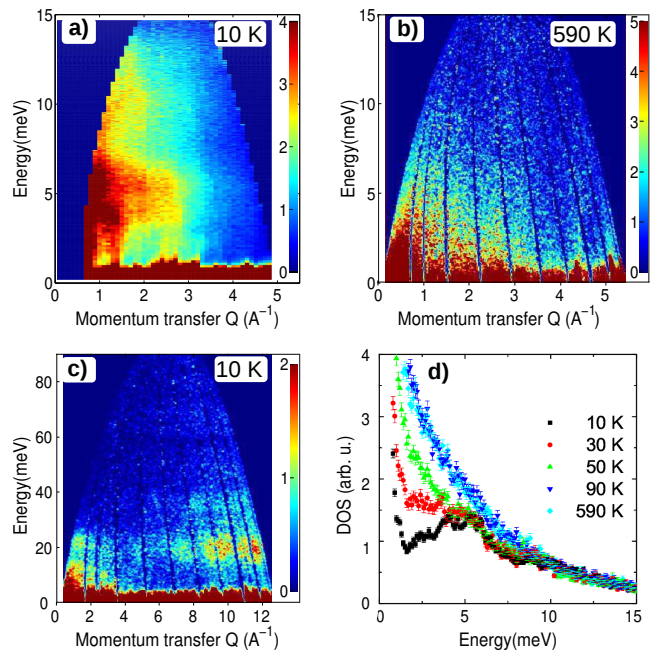}
  \caption{Bose-Einstein-factor-normalized INS intensity at (a) 10\,K and 
		  (b)
590\,K. (c) Map of INS intensity at 10\,K in broader \textit{Q} and
\textit{E} ranges than in (a), showing additionally the phonon absorption at 
higher \textit{Q}. Note the different
intensity scales in parts (a)--(c). (d) Magnon DOS determined by integrating the 
INS maps over the \textit{Q} range from 0.5 to 3\,$\mbox{\AA}^{-1}$. The data 
come from the ISIS facility except the part (a) which was measured in ILL.}
	  \label{fig:INS}
\end{figure}

In a polycrystalline sample, it is not possible to determine directly the 
	phonon and magnon dispersion branches in the BZ. Instead, the data obtained 
	by INS correspond to an orientation-averaged scattering function $S(Q,E)$ 
	where $Q$ is the total momentum transfer and $E$ the energy transferred from 
	the neutrons to the crystal lattice. Beyond the quasielastic neutron 
	scattering (QENS) region, delimited by the energy transfer of 
	$E\approx 1$\,meV, three scattering intensity maxima are observed at 10\,K for $Q<1\,\rm\AA^{-1}$ 
	(see Fig.~\ref{fig:INS}a); they occur near $E=3.5$, 6 and 10\,meV, in a very good 
	agreement with the frequencies of the modes observed in IR transmittance and 
	THz spectra (see Fig.~\ref{fig:lowFallT}).  It is worth noting that upon 
	heating, this scattering zone becomes broader but it can still be detected 
	up to $T=590\,\rm K$ (see Fig.~\ref{fig:INS}b). The magnetic character of 
	these excitations is unambiguously established
\cite{Shirane06} by the fact that the scattering intensity of magnons decreases 
with
increasing $Q$. In contrast, the phonons can be observed at higher values of 
$Q$; in our case, they were detected at $Q>7\,\rm\AA^{-1}$ (see 
Fig.~\ref{fig:INS}c).
By integrating $S(Q,E)$ over a low-$Q$ interval for different temperatures, 
	one can obtain the magnon DOS (see Fig.~\ref{fig:INS}d). 	 In 
		general, the peaks in the DOS correspond to extremes of magnon branches, 
		i.e., to magnons either from the BZ center, BZ boundary or to a minimum 
	of a magnon branch if the magnetic structure is modulated.  It is well known 
that electromagnons can be activated in the IR spectra just at these 
wavevectors \cite{Stenberg12}. Therefore, all the available evidence supports the hypothesis of 
electromagnon modes in our spectra. 

Furthermore, one can see a weaker magnon signal in the INS spectra peaking 
around 20\,meV and $Q<2\,\rm\AA^{-1}$ (see Fig.~\ref{fig:INS}c). We 
attribute this signal to two-magnon scattering, which corresponds to 
the observed maximum in the INS intensity at 10\,meV (see 
Fig.~\ref{fig:INS}a).  Alternatively, the 
weak 
magnon maximum could be due to one-magnon scattering; one can find also a matching vibrational mode near 20\,meV in the IR spectra
(Figs.~\ref{fig:IRrefl} and \ref{fig:mody-T}). However, its oscillator strength seems to 
be too high for a magnon and the shape of the reflectivity spectra does not 
correspond to a magnon \cite{Mukhin91}. Therefore, we conclude that the 
(electro)magnons probably exist up to about 10\,meV only.

\begin{figure}[b!]
	  \centering
	  \includegraphics[width=0.85\columnwidth]{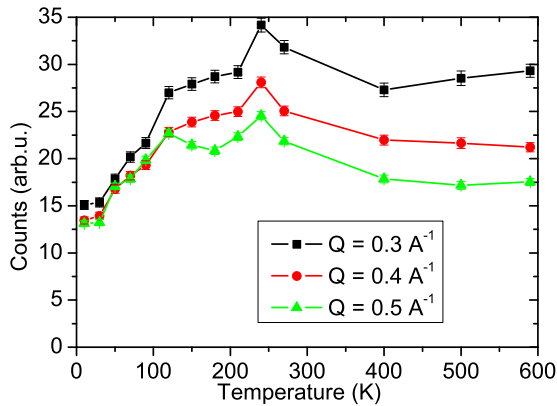}
	  \caption{Temperature dependence of the QENS as for selected values of $Q$.}
	  \label{fig:QNS-q}
\end{figure}

Whereas the existence of the lower-energy magnon band (near 3\,meV) in the 
IR-active spectra is clearly limited to the phase with long-range magnetic 
ordering which exists below $T_{\rm N2}$, the mode near 6\,meV is strong below 
$T_{\rm N1}$ but it weakly persists far above $T_{\rm N1}$ (see 
Fig.~\ref{fig:lowFallT}), as does the magnetic-field sensitive mode near 
9\,meV. These observations lead us to the conclusion that some 
kind of short-range magnetic ordering exists even if the long-range magnetic 
ordering is suppressed by the thermal vibrations. An analogous 
observation of short-range magnetic order, 
persisting 15\,K above the N\'eel temperature, was reported 
recently for non-stoichiometric multiferroic 
$\mbox{Ca}_{1-x}\mbox{Cu}_x\rm Mn_7O_{12}$\cite{Slawinski12} with $x\ge0.1$; to 
our knowledge, no such effect was observed yet for the stoichiometric compound 
with $x=0$. We suppose that the absorption 
maxima near 6 and 9\,meV might be linked to  short-wavelength 
(high-wavevector) magnon modes situated near the BZ boundary
and, since these modes are observed throughout the paramagnetic phase, we 
suggest to call them \textit{paraelectromagnons}. Indications of similar overdamped 
excitations were observed earlier in paramagnetic phases of other 
multiferroics, e.g.\ in $\rm TbMnO_3$ \cite{Takahashi08}, $\rm 
GdMnO_3$~\cite{Shuvaev11a}, $\rm TbMn_2O_5$ or $\rm 
YMn_2O_5$~\cite{Sushkov98}, but their origin and the reason for their 
activation in the THz spectra were not discussed in the literature yet. Based 
on our experimental results, we believe that the paraelectromagnon 
frequencies 
correspond to those of  magnons from the BZ boundary; 
their activation would 
be a consequence of a relaxation in selection rules in the incommensurately 
modulated magnetic structure. Further, we assume that the paraelectromagnons
exist far above $T_{\rm N1}$ due to short-range magnetic correlations and 
that they are excited by the electric component of the electromagnetic 
radiation due to magnetostriction, in the same way as zone-boundary 
electromagnons below $T_{\rm N}$.\cite{ValdesAguilar09,Stenberg12}

The idea of short-range magnetic ordering at temperatures far from the magnetic 
phase is further supported by our results of the QENS. Fig.~\ref{fig:QNS-q} 
shows the temperature dependence of the scattering intensity $S(Q,E)$ integrated 
over the energy interval $0\le E \le1$\,meV as a function of the momentum 
transfer $Q$.  Within the whole temperature range (10 to 590\,K), a contribution 
due to coherent QENS was observed with an intensity substantially exceeding the 
noise floor. Remarkably, upon heating above $T_{\rm N1}$, the QENS intensity 
increases up to $T_{\rm inc}$ where it reaches the maximum. On further heating, 
a decrease is observed, but the QENS intensity remains high and 
approximately constant even in the metallic cubic phase, up to 590\,K.

\begin{figure}
	  \centering
	  \includegraphics[width=0.9\columnwidth]{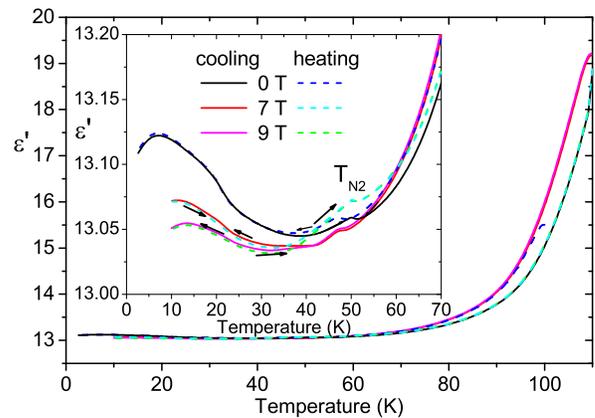}
	  \caption{Temperature dependence of the permittivity at 1\,kHz upon cooling 
	  and heating with various intensities of external magnetic field. The 
  inset shows in detail the temperature region below 70\,K.}
	  \label{fig:eps-B}
  \end{figure}

It is worth noting that our observation of presumed electromagnons differs 
from those reported in single-crystal multiferroics, where a step-like increase 
in the permittivity often occurs below the temperature where the electromagnon 
activates.~\cite{Sushkov08} In contrast, for our polycrystalline sample of $\rm 
CaMn_7O_{12}$, upon cooling, the permittivity decreases substantially down to 
40\,K due to a drop in conductivity (above 100\,K, the loss tangent exceeds 
unity); we suppose that this compensates the expected increase in the real part 
of the permittivity upon electromagnon activation. Nevertheless, we observed a 
small peak in $\varepsilon(T)$ at $T_{\rm N2}$, followed by an increase below 
40\,K (see Fig.~\ref{fig:eps-B}). One can also observe a decrease in 
permittivity with magnetic field, which can be explained by stiffening of the 
electromagnons.

\section{Conclusion}
We have performed a thorough spectroscopic study of polycrystalline multiferroic 
$\rm CaMn_7O_{12}$ over a wide range of temperatures and applied  
magnetic fields.  Our analysis of the FTIR, THz time-domain and INS 
spectra proves the existence of magnon modes, including probably one 
electromagnon at 6\,meV and one at 10\,meV.   
In the paramagnetic phase, these modes become heavily damped, largely lose their oscillator strengths 
and disappear as much as 100--200\,K 
above $T_{\rm N1}$. We propose to call such spin modes 
paraelectromagnons, by
analogy with paramagnons. Although our experimental results do not allow us to 
draw a precise microscopic picture of these modes, the absence of long-range 
magnetic ordering in the incommensurate phase lets us conclude that the observed 
absorption peaks near 6 and 10\,meV can correspond to dynamical short-range correlations 
which give rise to a maximum in the magnon DOS at the edge of the BZ. These peaks, 
together with the unusually strong QENS present up to high temperatures, 
are apparently linked to the underlying structure and to strong short-range 
spin-spin interactions, which are also at the origin of the record ferroelectric 
polarization in $\rm CaMn_7O_{12}$.

Even if our experiments show that the magnetoelectric character of the 
excitations seen below 10\,meV is highly probable, its final confirmation could 
be obtained only by measurements of directional dichroism or by a detailed 
polarization analysis of the THz spectra. Unfortunately, such experiments 
require relatively large single crystals of $\rm CaMn_7O_{12}$, which are 
currently not available.

\begin{acknowledgments}
This work was supported by the Czech Science Foundation (project No.\ 
P204/12/1163). The kHz-frequency experiments were performed in MLTL (see http://mltl.eu), which is supported within the          
program of Czech Research Infrastructures (project No.\ LM2011025).
The experiments in ILL Grenoble were carried out as a part of the project 
LG11024 financed by the Ministry of Education of the Czech Republic.  
\end{acknowledgments}

\end{document}